\documentclass[twocolumn]{aastex62}

\graphicspath{{./}{figures/}}

\begin{document}

\title{Deep Hubble Space Telescope Imaging on the Extended Ly$\alpha$ Emission of a QSO at $z = 2.19$ with Damped Lyman Alpha System as a Natural Coronagraph}

\author{Jiani Ding}
\affil{Department of Astronomy and Astrophysics, UCO/Lick Observatory, University of California, 1156 High Street,
Santa Cruz, CA 95064, USA}
\author{Zheng Cai}
\affil{Department of Astronomy and Center for Astrophysics, Tsinghua University, Beijing, China 100084; zcai@mail.tsinghua.edu.cn}
\author{J. Xavier Prochaska}
\affil{UCO/Lick Observatory, University of California, 1156 High Street, Santa Cruz, CA 95064, USA}
 \affil{Kavli Institute for the Physics and Mathematics of the Universe (Kavli IPMU; WPI), The University of Tokyo, 5-1-5 Kashiwanoha, Kashiwa, Chiba Prefecture 277-8583, Japan}

\author{H.Finley}
\affil{Université de Toulouse, UPS-OMP, 31400 Toulouse, France}
\affil{IRAP, Institut de Recherche en Astrophysique et Planétologie, CNRS, 14 avenue Édouard Belin, 31400 Toulouse, France}
\author{Xiaohui Fan}
\affil{Steward Observatory, University of Arizona, Tucson, AZ 85721, USA}
\author{Zheng Zheng}
\affil{Department of Physics $\&$ Astronomy, University of Utah, Salt Lake City, UT 84112-0830, USA}
\author{H. Fathivavsari}
\affil{Institut d’Astrophysique de Paris, Universite Paris 6-CNRS, UMR7095, 98bis Boulevard Arago, F-75014 Paris, France}
\author{P. Petitjean}
\affil{Institut d’Astrophysique de Paris, Universite Paris 6-CNRS, UMR7095, 98bis Boulevard Arago, F-75014 Paris, France}

%
%
%
%





\begin{abstract}
Recent observations suggest that proximate damped Ly$\alpha$ (PDLA) systems can be used to study the host galaxies of Quasi-stellar objects (QSOs), because the PDLAs can block the bright point-spread-function (PSF) from central QSOs. Using six-orbits of narrowband imaging with $\it{HST}$/WFC3, we present the first high resolution narrowband image of the Ly$\alpha$ emission in the PDLA trough of the QSO SDSSJ115432.67-021537.9 . We detect one major component and one minor component in the narrowband imaging. Combining the $\it{HST}$/WFC3 imaging with deep Magellan/MagE spectra, we measure that the Ly$\alpha$ flux F$_{\rm{Ly\alpha}}$ = 1.56$\pm0.10 \times10^{-16}$ $\rm{erg\ s^{-1}\ cm^{-2}}$, which is among the luminous ($\approx$ 2.7 L$^{\star}_{\rm{Ly\alpha}}$) Ly$\alpha$ emitters at $\it{z} =$ 2.19. The $\rm{Ly\alpha}$-based star formation rate (SFR) is $\sim$ 7 $\rm{M_{\sun} \ yr^{-1}}$. These observational results favor that 
 the star formation from the host galaxy could be the main mechanism to power the Ly$\alpha$ emission. This new method sheds new light on the study of the kinematic structure and the spatial distribution of the extended Ly$\alpha$ emitting regions around the QSO host.

\end{abstract}






\section{INTRODUCTION} 
\label{sec:style}
High redshift QSO host galaxies are crucial for studying the coevolution of massive galaxies and their central black holes 
(\citealt{Schramm2008,McLeod2009,Targett2012,Matsuoka2015}). The relation between black hole mass and bulge stellar mass connects the star formation history of the host galaxies with the evolution of the central black holes (e.g., \citealt{Gebhardt2000,McConnell2013,Bennert2015,Reines2015}).
Studying the host galaxies casts light on segregating the star formation process and the black hole growth, leading to a better understanding of galaxy evolution and provides strong tests of merger-driven evolutionary model (e.g., \citealt{Hopkins2006}). To better explore host galaxies around QSOs, a technique has been proposed to utilize proximity damped Ly$\alpha$ absorbers (PDLAs) at approximately the same redshift as the QSOs.
The DLA clouds are expected to cover the Broad Line Region of the QSO ($\sim$ pc scales) but leave the extended emission unshielded at greater than about 1 kpc scales. 
Therefore, DLAs can be used as natural coronagraphs, to reveal the extended 
emission from QSO host 
(e.g., \citealt{Hennawi2009,Finley2013,Cai2014,Fa2015,Fa2016,Fa2018}). 
With analysis of $\ge$ 2,000 high S/N DLAs in SDSS-III, \citet{Cai2014} indicate that due to the correlation between the residual intensities in the DLA troughs and the quasar
luminosities, the residual flux detected in the DLA trough are more likely from QSO hosts rather than the DLA galaxies. 


\citet{Finley2013} identified 26 $\it{z}$ $>$ 2 PDLAs that have narrow Ly$\alpha$ emission detected in the PDLA trough with a FWHM $\sim$ 500 $\rm{km \ s^{-1}}$. 
There are several possible origins for this extended emission, namely: Ly$\alpha$ emission from the star-forming regions in the QSO host galaxy, fluorescent recombination radiation powered by the QSO, or Ly$\alpha$ emission from the DLA host. 
\citet{Fa2016} conduct a detailed spectroscopic study on a sample of six QSOs with PDLA and suggest that three of the targets have luminosity of Ly$\alpha$ similar to bright LAEs  ($\rm{L = 2-4 \ L^*_{\rm{Ly\alpha}}}$), consistent with the expected 
Ly$\alpha$ emission from QSO host galaxy.



To further investigate this technique, we conduct a high-resolution, direct-imaging program. 
In this paper, we present a pilot imaging on J1154-0215, one PDLA at $z=2.1853$ reported by \citet{Finley2013} that has Ly$\alpha$ emission perfectly residing within the HST WFC3/UVIS FQ387N narrowband filter. 
Using six-orbits of high-resolution HST imaging and the deep spectroscopic measurement from the Magellan/MagE spectrum from \citet{Fa2016}, we discuss possible origins of the extended Ly$\alpha$ emission from J1154-0215. Organization of this Letter is as follows. In \S2, we discuss our HST observations and data reduction. In \S3, we measure the column density of the DLA and the Ly$\alpha$ flux of the target and discuss its morphology.
 In \S4, we discuss the basic physical properties of the DLA and possible physical scenarios for the origin of the Ly$\alpha$ emission. Throughout the whole paper, we adopt a flat lambda-CDM cosmology with $\rm{\Omega_{\lambda}=0.7,\ \Omega_{m}=0.3\ and\ H_{0}=70\ km\ s^{-1}\ Mpc^{-1}}$.
 
\section{OBSERVATION AND DATA REDUCTION} \label{sec:style}

\subsection{Magellan/MagE Spectroscopy}

Fig.\ref{tab:figure1} shows the spectroscopic data of J1154-0215 from \citet{Fa2016}.
The median signal-to-noise (S/N) ratio around the Ly$\alpha$ emission is $\sim$ 3 /pixel and the resolution is (R $\sim$ 4000). The detailed reduction and data analysis process for the final spectrum can be found in \citet{Fa2016}. 
The MagE spectroscopy allows us to measure the metal lines associated with this PDLA. The metal line diagnostic will constrain the distance between the absorbers and the QSO (See details in \S 4.1). 

\subsection{HST Narrowband Imaging}
We use six orbits (about 17,000s) on the FQ387N narrowband observation of the QSO J1145-0215, which allows us to detect the Ly$\alpha$ flux at 15-$\sigma$ level in the FQ387N filter.  

\indent We distribute our entire observations into two individual visits, each with three orbits. A standard three-point dither sequence is applied to populate each orbit. 
\indent We conduct the deep HST/WFC3 narrowband FQ387N imaging of J1154-0215 as explained above. The redshift for the Ly$\alpha$ emission $\it{z_{em}}$ = 2.181 and the redshift for the absorber, $\it{z_{\rm{abs}}}$ = 2.1853, which are measured in \citet{Fa2016}.
In Fig. \ref{tab:figure1}, we plot the spectrum with filter response curve of the FQ387N filter   (${\lambda_{c} = 3874\ \rm{\AA}}$, 
FWHM ${= 34\ \rm{\AA}}$) and the Voigt profile fitting for the DLA.  The filter lies within the DLA trough and the Ly$\alpha$ emission is included in the filter. 

The data reduction is conducted using WFC3/UVIS \citep{Koekemoer2002}, and the detailed procedures follow \citet{Cai2011,Cai2015}. To optimize the output data quality, we choose a final output pixel scale of 0.03$"$ instead of initial pixel scale 0.04$"$ and final pixfrac parameter 0.7 (shrinking pixel area) after different trials of combinations of parameters. The final output images we obtained from HST for J1145-0215 is shown in the upper left panel of Fig. \ref{tab:figure2}.

\section{RESULTS} \label{sec:floathttps://www.overleaf.com/15185355nhmbtctcmhfcs}

\subsection{The Detection of Extended Ly$\alpha$ Emission by HST}

In our HST narrowband image, we detect spatially resolved, diffuse Ly$\alpha$ emission. 
Using SExtractor \citep{Bertin1996}, we measure the Ly$\alpha$ flux using an elliptical aperture with a Kron factor of 2.5 ($0.075"$) (magenta elliptical in upper left panel in Fig. \ref{tab:figure2}). The flux density of the source is F$_{\rm{FQ387N}}$ = $6.2\pm0.2 \times10^{-18}\ \mathrm{erg\ s^{-1}\ cm^{-2} \AA^{-1} }$ (total flux F$_{\rm{total}}$ = $2.11\pm0.07 \times10^{-16}$ $\mathrm{erg\ s^{-1}\ cm^{-2}}$), corresponding to an AB magnitude of $ 22.68 \pm 0.04$. 


In Figure \ref{tab:figure2} upper left panel, we overplot QSO position (dark point) with its error (green dashed circle).  Note that our HST narrowband completely resides in the DLA dark trough. Thus, we do not have QSO continuum flux from 
our HST observations. The SDSS provides the QSO continuum position. 
We use two stars in the HST field to correct the offset between the SDSS and the HST coordinate systems (assuming the rotational offset is zero). 
We find that the offset between HST and SDSS is 0.09 arcsec +/- 0.1 arcsec. 
We also examine the morphology of this source. The HST image and the circularly averaged surface brightness (SB) profile shown in the lower left panel of Fig. \ref{tab:figure2} suggest that we detect two resolved components, and we use two Sersic components (right in Fig. \ref{tab:figure2}) to fit J1154-0215. 
We measure an effective radius of 0.29$\pm$0.05 kpc for the major component and of 1.52$\pm$0.6 kpc for the minor component. The contour of the SB of the source from the Sersic components model fitting is shown in the upper right panel of Fig. \ref{tab:figure2}. The residual of the fitting for this model is shown in the lower right panel of Fig. \ref{tab:figure2}.
We also use two Gaussian components to repeat the above fitting and recover similar results of one major component and one minor component. The detailed parameters of the two models are shown in table \ref{tab:table1}. 


\subsection{DLA column density and Ly$\alpha$ emission flux from MagE spectrum}

For the MagE spectra from \citet{Fa2016}, we plot the spectra with filter response curve of the FQ387N filter and the Voigt profile fitting with error for the DLA in figure \ref{tab:figure1}. The spectral resolution is 75 km/s. 
We attain the $\rm{H\,{\footnotesize I}}$ column density for the DLA log(N$_{\rm{HI}}$) = 21.76$\pm$0.1, consistent with the value fitted by \citet{Fa2016}. We further fit a double Gaussian profile to the Ly$\alpha$ spectrum inside the DLA dark trough. The integrated Ly$\alpha$ emission flux is F$_{\rm{Ly\alpha}}$ = 1.56$\pm0.10 \times10^{-16} \  \rm{erg\ s^{-1}\ cm^{-2}}$  from this spectrum. 
Moreover, we also detect and fit metal lines Si II1260, Si$^{*}$ II1264 in this spectrum (showed in Fig. \ref{tab:figure3} and detailed analysis in \S 4.1.). 

\section{DISCUSSION} \label{sec:floats}

In the following sub-sections, we discuss the physical properties of the DLA gas and the origin of the extended Ly$\alpha$ emission. 


\subsection{Size of the cloud and the distance to nearby QSO}

From the high resolution HST FQ387N imaging, we do not detect Ly$\alpha$ emission in the central QSO region, and we now refer to
this region as a Ly$\alpha$ ``void". This may be explained by the fact that a kpc scale optically thick cloud is located in the region of the Ly$\alpha$ emission ``void". We construct a power law model without DLA absorption for the SB profile centered at the quasar position (labeled in red in Fig. \ref{tab:figure5}). We then add the absorption from the DLA to the model and find the best-fitted model to the data of the SB profile \footnote{We assume the optical depth $\rm{\tau}$ of the target evolves as a function of the distance r ($\rm{\tau}$ = a$_{1}$*r$^{(-\alpha)}$) from the quasar position and the SB profile evolve as a$_{2}$*e$^{\rm{-\tau}}$+c, where a$_{1}$, $\alpha$, a$_{2}$ and c are constants.} (model labeled in green and data labeled in blue in Fig. \ref{tab:figure5}). The two models (with and without DLA implemented) converge at r $\sim$ 1.5 kpc. We thus infer that the cloud size optically thick to the Ly$\alpha$ emission is $\sim$ 1.5 kpc. Our conclusion also agrees with the physical extent ($<$ 10 kpc) of log(N$_{\rm{HI}}$) $\ge$ 20.3 absorption constrained by lensed quasar pairs in \citet{Cooke2010}. The physical picture of the QSO, DLA cloud and the ``void" of Ly$\alpha$ emission are illustrated in the left panel of Fig. \ref{tab:figure4}.  


From the last section, we also detect absorption lines, Si~II\,1260 and Si$^{*}$~II\,1264 from the MagE spectrum arising from the ground state n and the excited state $\rm{n^{*}}$ of Si$^{+}$. For a single, homogeneous cloud, the column density ratio between the lines from the excited state and the lines from the ground state is equal to the corresponding volume densities ratio of the ions/atoms \citep{Bahcall1967, Silva2002, Prochaska2006}, namely, $\rm{\frac{n^{*}}{n}}$ = $\rm{\frac{\it{N}^{*}}{\it{N}}}$. This ratio may be determined by the ratio of the column density between Si$^{*}$~II\,1264 and Si~II\,1260. Since the Si~II\,1260 line is saturated, we will put a lower limit on its column density. Since the $b$ value and column density will be degenerated when fitting the line Si~II\,1260, we test a range of Doppler parameter (b) value and use the maximum b value (b = 30 km s$^{-1}$) to obtain a lower limit of the column density of $\rm{log(N/cm^{-2})}$ = 15.3 for Si II1260. We then fit Si$^{*}$ II1264  with a fixed b value = 30 km s$^{-1}$ and attain the column density of $\rm{log(N/cm^{-2})}$ = 13.62$\pm$0.13 \footnote{Since the S/N of the spectrum is low in the region these absorption lines located, we fit a Voigt profile for a fixed b value by using $\it{curvefit}$ in scipy package. The 1-$\sigma$ error of the column density is estimated from the fitting error. These lines vary within the reported uncertainty when b changes from 20-30 km s$^{-1}$.}.  The results of the Voigt profile fitting of these absorption lines are shown in Fig. \ref{tab:figure3}. 

If we assume the excited level is populated by indirect pumping of 
far-UV photons, \citet{Prochaska2006} derive a relation between $\rm{\frac{n^{*}}{n}}$ (see Figure 7 in \citet{Prochaska2006}) and the far-UV flux. We can derive that d = $\sqrt{\frac{L}{\rm{F_{uv}}4\pi}}$, where L is the quasar luminosity in the far UV range (8 eV $<$ h$\nu$  $< $13.6 eV), and ${\rm{F_{\rm{uv}}}}$ is the far UV flux determined from $\rm{\frac{n^{*}}{n}}$ based on the relation of population ratio and intensity from \citet{Prochaska2006}. The quasar luminosity L is computed by integrating a local power law through the far UV frequency range ($\int a\rm{\nu}^{-\alpha_{Q}} d\rm{\nu}$), where a is a constant. The local power law in this integral is estimated by the SDSS u band and SDSS g band photometric result. We finally get L = 1.33$\times$10$^{45}$ erg $\rm{s}^{-1}$. We infer ${\rm{I_{\rm{uv}}}}$ = 32 erg $\rm{s^{-1}cm^{-2}}$ from the ratio of the column density estimated by the pair of Si$^{+}$ lines. Based on the equation d = $\rm{\sqrt{\frac{L}{F_{\rm{uv}}4\pi}}}$, we finally get a upper limit of d $\leq$ 0.6 kpc with the lower limit on the column density of Si II1260, which is consistent with the fact that this target has the highest Al III/Si II ratio reported for a PDLA in \citet{Fa2016}, indicating that it may be close to the AGN. The physical picture of the QSO and DLA position is illustrated in the right panel of Fig. \ref{tab:figure4}

\subsection{The origin of the emitting gas}
Combining the photometric and $\it{Galfit}$ fitting results of the final combined image with the reduced 1-D spectrum, there are three main possible origins of the emitting gas for the Ly$\alpha$ emission: 1. Emission from star-forming regions in the QSO host galaxy. 2. Emission from star-forming regions in the DLA galaxy. 3. Fluorescent Ly$\alpha$ emission from the optically thick gas cloud. 

\subsubsection{QSO host galaxies} 
We compute the luminosity of the Ly$\alpha$ emission for target J1154-0215 to be $5.71\pm0.33$ $\times10^{42} \ \rm{erg \ s^{-1}}$. Most of the LAEs discussed in \citet{Ciardullo2012,Ouchi2008} have a Ly$\alpha$ emission $<$ 5$\times10^{43} \ \rm{erg \ s^{-1}}$, suggesting that the Ly$\alpha$ emission from this target may have similar luminosity as bright LAEs ($\sim$ 2.7 $\rm{L^{\star}_{\rm{Ly\alpha}}}$). If we assume the Ly$\alpha$ emission is from the star forming region of the host galaxy, we can estimate that the Ly$\alpha$-based star formation rate is  $\sim$ 7 M$_{\sun}$ yr$^{-1}$ (e.g., \citealt{Dijkstra2010,Kennicutt1998}).  
Taking into account the effect of dust extinction and absorption by foreground $\rm{H\, \small{I}}$ gas, the SFR we computed may be a lower limit. 

Assuming the Ly$\alpha$ emission originated from star-forming regions in the QSO host galaxy, the total fluxes F$_{\rm{total}}$ we measured inside the filter will be the sum of the UV continuum and the Ly$\alpha$ emission. Thus, we can estimate the UV continuum by subtracting the Ly$\alpha$ emission fluxes fitted by the spectrum from F$_{\rm{total}}$. The resultant UV continuum is F$_{\rm{UV}}$ = 0.86$\pm 0.3 \times 10^{-16} \ \rm{erg\ s^{-1}\ cm^{-2}}$, which is at the level of 14$\times$10$^{28}$ $\mathrm{erg\ s^{-1}\ Hz^{-1}}$. Assuming a flat UV-slope, the UV-based SFR is $\sim$ 19 M$_{\sun}$ yr$^{-1}$ \citep{Kennicutt1998}, which is consistent with the SFR inferred by the Ly$\alpha$ emission\footnote{SFR$_{\rm{UV}}$ = $\frac{f_{1500} \times 4 \rm{\pi} d^{2}_{L}  \times 1.4 \times 10^{-28}}{1+\it{z}}$ $\rm{M_{\sun} \ yr^{-1}}$ \citep{Kennicutt1998}}. The SFR$_{\rm{UV}}$ is consistent with the results of $\sim$ 200 LAEs at $\it{z}$ $\sim$ 2  \citep{Ciardullo2012,Gu2010,Ouchi2008}. In addition, \citet{Fa2016} suggest that this target exhibits typical features of LBGs (redshifted emission and blueshifted absorption). 
Also, the Ly$\alpha$ emission luminosity for our target do not exceed the luminosity of the brightest LBGs in \citet{Shapley2003}. Therefore, it is likely that the emission is coming from star formation based on these calculations.

The asymmetric morphology of the target fitted by $\it{Galfit}$ shows a major component and a minor component (see Fig. \ref{tab:figure2}). 
From the fitting models, we can see that the flux from the major component is twice of the fluxes from the minor component.  From the fitting result, the effective radius of the major component is approximately five times smaller than the minor component, suggesting that it is much more compact than the minor component. The simulation result of minor merger from \citet{Lotz2010} also suggests that the remnants of minor merger will show compact nuclei, which is consistent with our morphology results. Our current data may suggest that this QSO is associated with a merger event. 

\subsubsection{DLA galaxies} 
From \citet{Fa2016}, the metallicity of the associated DLA for J1154-0215 is [Si II/ H] = -1.70. Based on the relation between metallicity and stellar mass inferred in \citet{Moller2013} and relation between stellar mass and star formation rate demonstrated in \citet{Yates2012} (assuming the DLA system as a DLA galaxy), this low metallicity may correspond to a much lower star formation rate ($<$ 1 M$_{\sun}$ yr$^{-1}$) compared to the SFR estimated from Ly$\alpha$ emission and UV emission. The Ly$\alpha$ emission and the low metallicity DLA may be from separate system, suggesting that the Ly$\alpha$ emission is not coming from the DLA galaxy. 

\subsubsection{Fluorescent Ly$\alpha$ emission}
The third possibility for the origin of the Ly$\alpha$ emission is fluorescent Ly$\alpha$ emission from the optically thick gas that is close to the DLA cloud or from a system detached from it. Ionizing photons intercepted by the gas ionize neutral hydrogen atoms and the subsequent recombination has a high probability ($\eta \sim 0.6$; e.g., \citealt{GouldWeinberg1996}) of ending up as Ly$\alpha$ photons. 

We can estimate the surface brightness $I_\alpha$ of the fluorescent Ly$\alpha$ emission through 
$\pi I_\alpha = h\nu_\alpha \eta f F_i / (1+z)^4$, where $h\nu_\alpha$ is the Ly$\alpha$ photon energy and the $(1+z)^4$ factor accounts for the cosmological dimming. For isotropic quasar emission, the ionizing photon flux $F_i$ at the cloud position is $F_i = \int L_\nu/(h\nu) d\nu / (4\pi d^2)$, with $L_\nu$ the quasar luminosity and $d$ the distance from the quasar to the gas cloud. The integral goes from the Lyman limit $\nu_L=13.6\, {\rm eV}/h$ to infinity. The factor $f$ is introduced to scale the line-of-sight flux to that along the quasar-cloud direction, accounting for the anisotropy of the quasar emission of ionizing photons. With the assumption that $L_\nu = L_{\nu_L} (\nu/\nu_L)^{-\alpha}$, we have
\begin{equation}
I_\alpha = \frac{1}{(1+z)^4}\frac{1}{4\pi^2 d^2} \frac{\eta f}{\alpha} \frac{\nu_\alpha}{\nu_L} \nu_L L_{\nu_L}.
\end{equation}
The quasar has $\nu_L L_{\nu_L} = 0.76\times 10^{45} {\rm erg\, s^{-1}}$ with $\alpha=1.57$.
For $d\sim 0.6$ kpc and $f\sim 0.1$, we obtain the Ly$\alpha$ surface brightness to be at the level of
$0.8\times 10^{-13\,} {\rm erg\, s^{-1} cm^{-2} arcsec^{-2}}$, which is about two orders of magnitude higher than the observed surface brightness shown in Fig. \ref{tab:figure2}. To match the observation, we need to put the cloud much farther way from the quasar (e.g., $d >10$ kpc along the line of sight) or to have extremely anisotropic emission of the quasar (e.g., $f<10^{-3}$). 

The other difficulty faced by the fluorescent origin of the Ly$\alpha$ emission is the shape of the spectrum. Fluorescent Ly$\alpha$ photons emerging from the thin skin layer of the optically-thick cloud would typically have a double-peak profile (e.g., \citealt{GouldWeinberg1996, Zheng2002, Adelberger2006}), with peak separation about eight times the velocity dispersion in the cloud. If the velocity dispersion in the cloud is of the order of $50 {\rm km\, s^{-1}}$ (e.g., \citealt{Adelberger2006}), the two peaks (with separation $\sim 400 {\rm km\, s^{-1}}$) would be well resolved, which does not seem to be the case in the spectrum shown in Fig. \ref{tab:figure1}. The tentatively detected blue peak with low flux could indicate gas kinematics, like outflowing, in the cloud.

Overall, based on the Ly$\alpha$ surface brightness and the Ly$\alpha$ line profile, we find that the fluorescent origin of the Ly$\alpha$ emission is not favored.


%


\section{CONCLUSION and FUTURE OBSERVATION}
\begin{figure*}[ht!]
\centering
\figurenum{1}
\includegraphics[width=190mm]{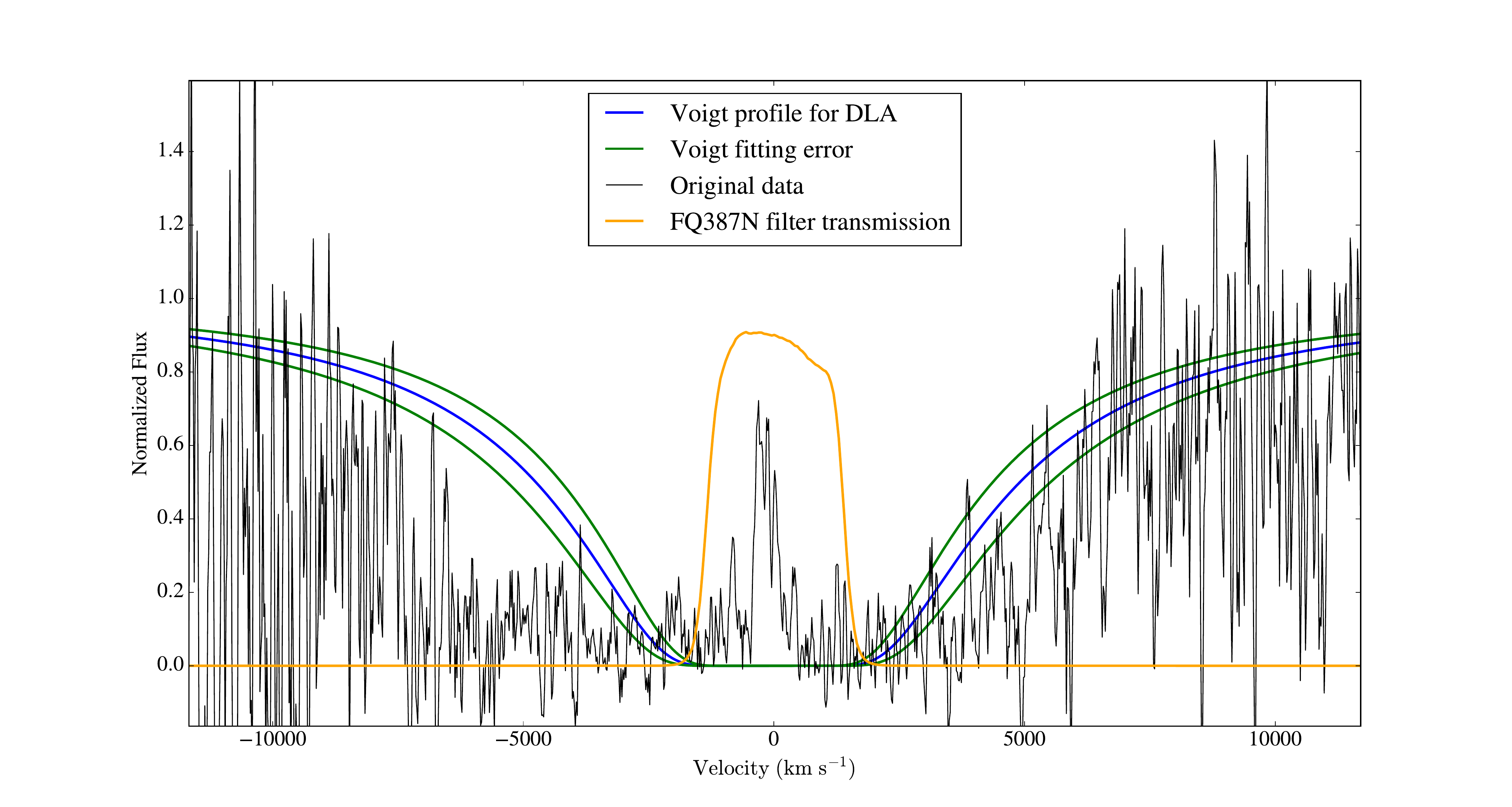}
\caption{The Voigt fitting (light blue) with error (green) for the target J1154-0215. The filter response (orange curve) for WFC3 FQ387N filter and original data (black curve) are also over-plotted in the graph. The absorption redshift for this target is $\it{z}$ =2.1853. \label{tab:figure1}}
\end{figure*}

\begin{figure*}[ht!]
\centering
\figurenum{2}
\includegraphics[width=190mm]{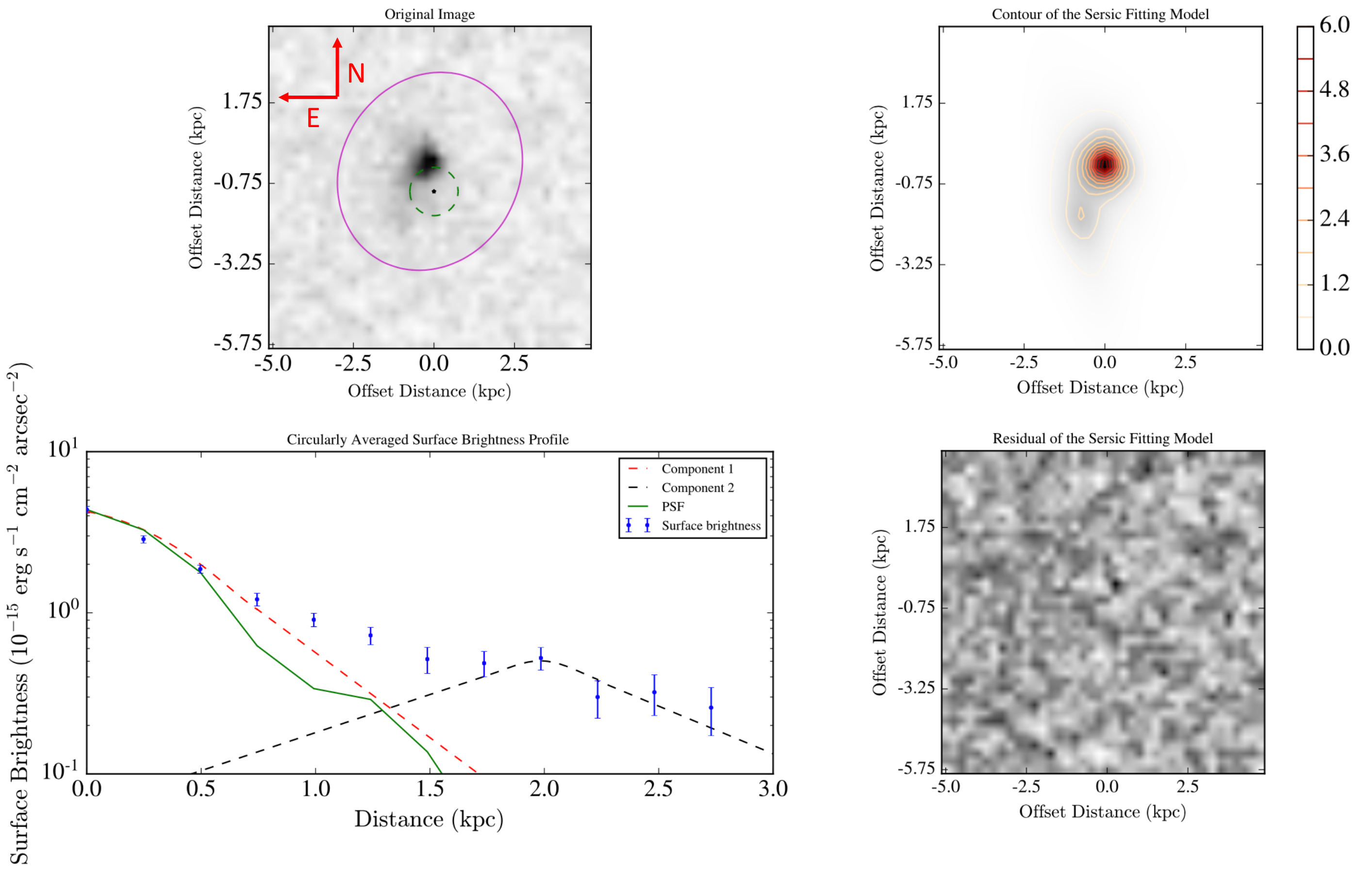}
\caption{The final combined image and model of the source. Top-left panel: The final combined image of the source. The aperture we used for photometric measurements is labeled in magenta. The position of the QSO is labeled in black dot and the error of the position is labeled in green. Bottom-left panel: The circularly averaged surface brightness (SB) profile (data points labeled in blue) centered at the flux peak of the source. The PSF profile is labeled in green curve and re-normalized to set the first data point of PSF profile and the profile for the source to have equal flux. The red and black dashed curves show the contributions from the detected two components for the source. Upper Right Panel: The Galfit model of the source for fitting two Sersic components with the contours and colorbar of the SB of this model. The value of the colorbar is in the units of 10$^{-15}$ erg s$^{-1}$ cm$^{-2}$ arcsec$^{-2}$.  Lower Right Panel: The residual of the Galfit model for fitting with two Sersic components. The offset distances labeled in the axis of the images are offset distances from the flux peak of the target. \label{tab:figure2}}

\end{figure*}

\begin{figure*}[ht!]
\centering
\figurenum{3}
\includegraphics[width=200mm]{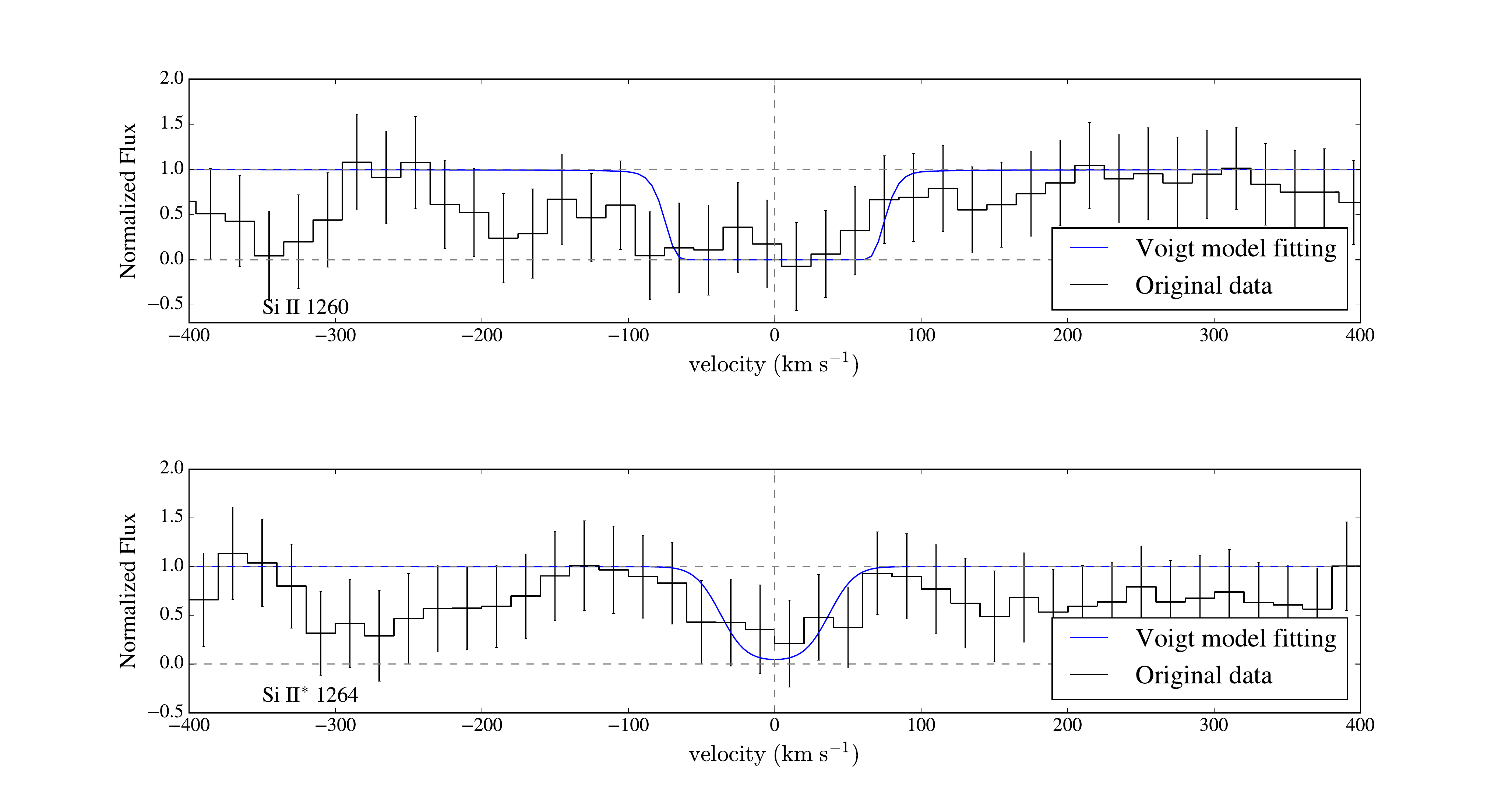} 
\caption{Velocity plots and best fit Voigt profile of the species detected in the DLA redshift ($\it{z}$ = 2.18532). The original data is labeled in black with error bar and the Voigt profile is labeled in blue.  \label{tab:figure3}}
\end{figure*}

\begin{figure*}[ht!]
\centering
\figurenum{4}
\includegraphics[width=180mm]{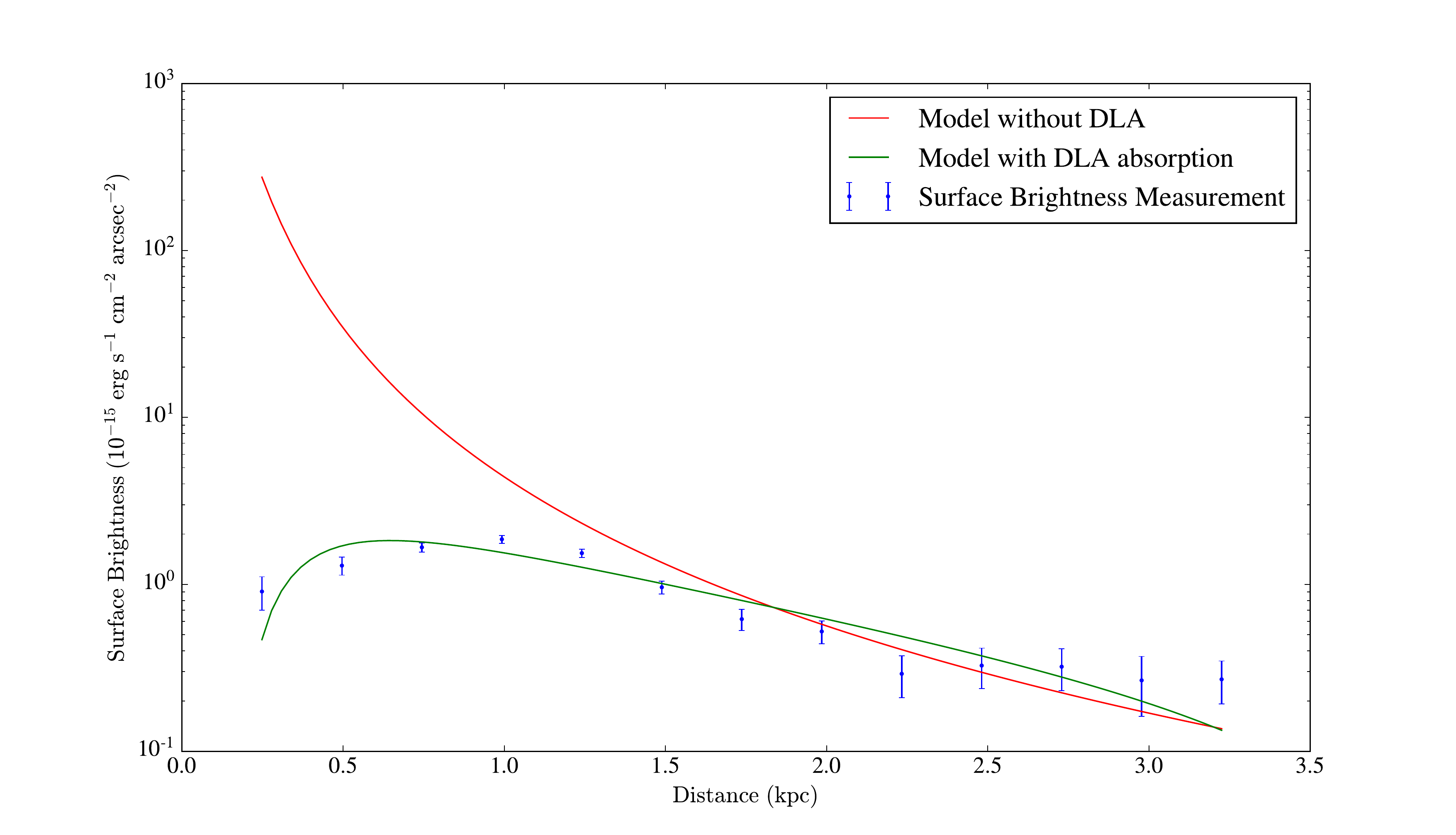}
\caption{Surface brightness profile and models for the profile with and without DLA absorption around the quasar (centering at the quasar position). The red curve is the model for the SB profile centering at the position of quasar without DLA absorption while the green curve is the model with DLA absorption. The model labeled in the green curve is the best fitted model to the data of the SB measurement centering at the center of the quasar (labeled in blue). \label{tab:figure5}}
\end{figure*}

\begin{figure*}[ht!]
\centering
\figurenum{5}
\includegraphics[width=180mm]{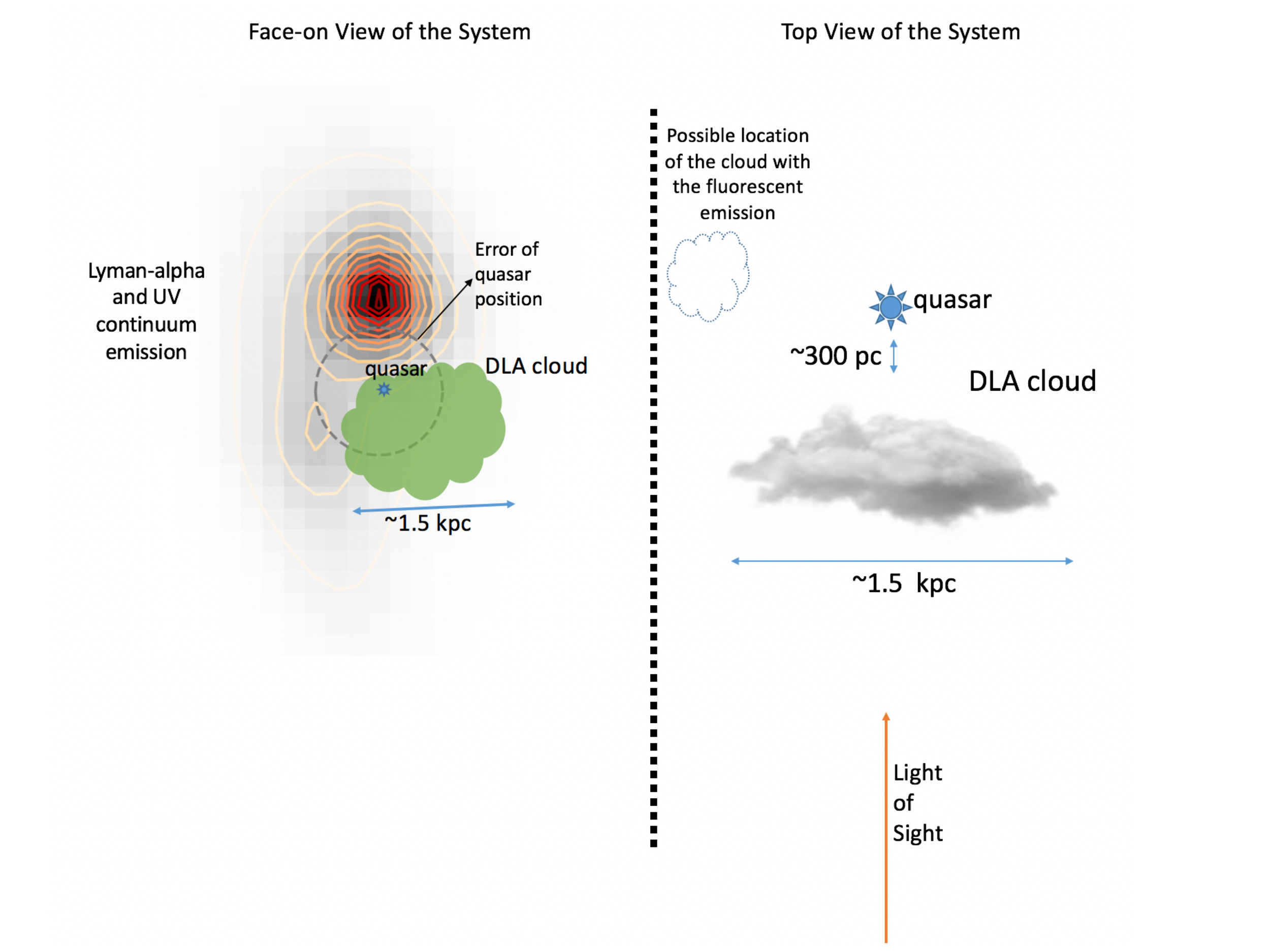}
\caption{Cartoon illustration of the physical picture of the J1154-0215 system. Left panel: Face-on view of the system. The contour plot is from the contour of the SB of the Sersic fitting model. The DLA cloud is labeled in green with its size. The QSO position is labeled in blue with its 1-$\rm{\sigma}$ error labeled in grey. Right panel: Top view of the system. The QSO is labeled in blue and the DLA cloud is labeled in white with its size and distance from the QSO. The cloud in dashed line shows the possible location of the gas cloud with the fluorescent emission for the less possible scenario that the fluorescent emission is the origin of the Ly$\alpha$ emission. \label{tab:figure4}}
\end{figure*}

Combing six-orbits of $\it{HST}$ WFC3/UVIS FQ387N filter deep imaging with the spectrum (Fig. \ref {tab:figure1}), we conduct a pilot study on a target J1154-0215 with a PDLA system, i.e. J1154-0214. The narrow Ly$\alpha$ emission is revealed inside the DLA dark trough. We prefer the conclusion that this Ly$\alpha$ emission may come from the star forming region of the QSO host galaxies. Future observation of the H-$\beta$ and other Balmer lines in the deep near-IR spectroscopy can help constrain the dust extinction. Furthermore, future IFU observations can help us to understand the gas kinematic structure of this system and thus further understand the interrelation between the central QSO, QSO host galaxies and galactic environment of this target. In order to further confirm the origin of the Ly$\alpha$ emission, it is crucial to understand the polarization of the emission. Future facilities, including Giant Magellan Telescope (GMT) and Thirty Meter Telescope (TMT) will probe the polarization of these targets with reasonable exposure time and thoroughly distinguish the origin of the emission, shedding new light on studying the extended Ly$\alpha$ emission from star forming region or fluorescent emission from gases around QSO. \\

We dedicate this article in memory of Hayley Finley who left
us a few days before the article was accepted. \\
We wish to thank the anonymous referee for their insightful comments that improved the paper. We also thank the support from NASA through grant
HST-GO-13818 from the Space Telescope Science Institute. We are grateful to acknowledge Luming Sun and Ning Jiang for sharing their insights on the future work of this paper. 

\newpage
\floattable
\begin{deluxetable}{ccccccc}
\tablecaption{Parameters of the two Galfit models for the source \label{tab:table1}}
\tablecolumns{7}
\tablenum{1}
\tablewidth{0pt}
\tablehead{ 
\colhead{model name} &
\colhead{comp1 mag} &
\colhead{comp2 mag} &
\colhead{comp1 effective radius} &
\colhead{comp1 Sersic index} &
\colhead{comp2 effective radius}&
\colhead{comp2 Sersic index}
 \\
\colhead{} &
\colhead{} &
\colhead{} &
\colhead{(kpc)} &
\colhead{} &
\colhead{(kpc)} &
\colhead{} 
}
\startdata
Sersic model & 23.02$\pm$0.07 & 23.82$\pm$0.28 & 0.29$\pm$0.05 & 4.43$\pm$1.95  & 1.52$\pm$0.6 & $2.49\pm1.28$    \\ 
Gaussian model & 23.22$\pm$0.02 & 23.80$\pm$0.07 & 0.62$\pm$0.06 & 0.28$\pm$0.12 & 2.97$\pm$0.33 & $0.23\pm0.05$   \\ 
\enddata
\tablecomments{1. The magnitude is AB magnitude. 2. The notation 'comp' means 'component'.}
\end{deluxetable}
\vspace{5mm}

\bibliographystyle{aasjournal.bst}
\bibliography{refs}

\allauthors

\listofchanges

\end{document}